\RequirePackage[hyphens]{url}
\newif\ifarxiv
\arxivtrue %
\ifarxiv
\documentclass[conference]{IEEEtran}
\usepackage{lmodern}
\usepackage{bm}
\usepackage{amsmath}
\usepackage{amssymb}
\else
\documentclass[sigconf]{acmart}

\makeatletter

\newcommand\addauthornote[1]{%
  \if@ACM@anonymous\else
    \g@addto@macro\addresses{\@addauthornotemark{#1}}%
  \fi}

\newcommand\@addauthornotemark[1]{\let\@tmpcnta\c@footnote
   \setcounter{footnote}{#1}\addtocounter{footnote}{-1}
    \g@addto@macro\@currentauthors{\footnotemark\relax\let\c@footnote\@tmpcnta}}

\makeatother
\fi
\usepackage{url}
\usepackage{xcolor}

\usepackage{booktabs}
\usepackage{graphicx}
\usepackage{subcaption}
\usepackage{algorithm}
\usepackage{algcompatible}

\usepackage{xy}
\usepackage[braket, qm]{qcircuit}

    \newcommand{\ketbra}[2]{\mathinner{|{#1}\rangle\langle{#2}|}}
    
    \newcommand{\Gate}[1]{\textsc{#1}}

    \newcommand{\ygate}{\Gate{y}}
    \newcommand{\xgate}{\Gate{x}}

    \newcommand{\idgate}{\Gate{i}}

    \newcommand{\iu}{\mathrm{i}\mkern1mu}  %

    \usepackage{listings}
    \usepackage{color}
    \usepackage{courier}
    \usepackage{hyperref}
    \usepackage{tikz}
    \usetikzlibrary{shadows.blur,arrows.meta,positioning,bending,fit}

    \definecolor{UBCblue}{HTML}{0578c3}
    \definecolor{UBCblueLight}{HTML}{86e2fd}

\ifarxiv
    \usepackage[T1]{fontenc}
    \usepackage{inconsolata}
\fi

    \definecolor{darkred}{rgb}{0.6,0.0,0.0}
    \definecolor{darkgreen}{rgb}{0,0.50,0}
    \definecolor{lightblue}{rgb}{0.0,0.42,0.91}
    \definecolor{orange}{rgb}{0.99,0.48,0.13}
    \definecolor{grass}{rgb}{0.18,0.80,0.18}
    \definecolor{pink}{rgb}{0.97,0.15,0.45}

    \lstdefinestyle{colored}{ %
      basicstyle=\ttfamily,
      backgroundcolor=\color{white},
      commentstyle=\colorn{green}\itshape,
      keywordstyle=\color{blue}\bfseries\itshape,
      stringstyle=\color{red},
    }
    \lstdefinelanguage{PythonPlus}[]{Python}{
      morekeywords=[1]{,as,assert,nonlocal,with,yield,self,True,False,None,} %
      morekeywords=[2]{,__init__,__add__,__mul__,__div__,__sub__,__call__,__getitem__,__setitem__,__eq__,__ne__,__nonzero__,__rmul__,__radd__,__repr__,__str__,__get__,__truediv__,__pow__,__name__,__future__,__all__,}, %
      morekeywords=[3]{,object,type,isinstance,copy,deepcopy,zip,enumerate,reversed,list,set,len,dict,tuple,range,xrange,append,execfile,real,imag,reduce,str,repr,}, %
      morekeywords=[4]{,Exception,NameError,IndexError,SyntaxError,TypeError,ValueError,OverflowError,ZeroDivisionError,}, %
      morekeywords=[5]{,ode,fsolve,sqrt,exp,sin,cos,arctan,arctan2,arccos,pi, array,norm,solve,dot,arange,isscalar,max,sum,flatten,shape,reshape,find,any,all,abs,plot,linspace,legend,quad,polyval,polyfit,hstack,concatenate,vstack,column_stack,empty,zeros,ones,rand,vander,grid,pcolor,eig,eigs,eigvals,svd,qr,tan,det,logspace,roll,min,mean,cumsum,cumprod,diff,vectorize,lstsq,cla,eye,xlabel,ylabel,squeeze,}, %
    }
    \lstdefinelanguage{PyBrIM}[]{PythonPlus}{
      emph={d,E,a,Fc28,Fy,Fu,D,des,supplier,Material,Rectangle,PyElmt},
    }
    \lstdefinestyle{colorEX}{
      basicstyle=\small\ttfamily,
      backgroundcolor=\color{white},
      commentstyle=\color{darkgreen}\slshape,
      keywordstyle=\color{blue}\bfseries\itshape,
      keywordstyle=[2]\color{blue}\bfseries,
      keywordstyle=[3]\color{grass},
      keywordstyle=[4]\color{red},
      keywordstyle=[5]\color{orange},
      stringstyle=\color{darkred},
      emphstyle=\color{pink}\underbar,
    }
    
    \newcommand{\qokit}{\texttt{QOKit}}

    \newenvironment{tolerant}[1]{%
      \par\tolerance=#1\relax
    }{%
      \par
    }

\begin{document}

\title{Fast Simulation of High-Depth QAOA Circuits}

\ifarxiv
\author{Danylo Lykov$^{\dagger,\ddagger}$, Ruslan Shaydulin$^{\dagger}$, Yue Sun$^{\dagger}$, Yuri Alexeev$^{\ddagger}$ and Marco Pistoia$^{\dagger}$ \\ 
$^{\dagger}${\small Global Technology Applied Research, JPMorgan Chase, New York, NY 10017, USA} \\
$^{\ddagger}${\small Computational Science Division, Argonne National Laboratory, Lemont, IL 60439, USA}}
\fi

\ifarxiv
\maketitle
\fi

\begin{abstract}
Until high-fidelity quantum computers with a large number of qubits become widely available, classical simulation remains a vital tool for algorithm design, tuning, and validation. We present a simulator for the Quantum Approximate Optimization Algorithm (QAOA). Our simulator is designed with the goal of reducing the computational cost of QAOA parameter optimization and supports both CPU and GPU execution. Our central observation is that the computational cost of both simulating the QAOA state and computing the QAOA objective to be optimized can be reduced by precomputing the diagonal Hamiltonian encoding the problem. We reduce the time for a typical QAOA parameter optimization by eleven times for $n = 26$ qubits compared to a state-of-the-art GPU quantum circuit simulator based on cuQuantum. Our simulator is available on GitHub: \url{https://github.com/jpmorganchase/QOKit}
\end{abstract}

\section{Introduction}

Quantum computers offer the prospect of accelerating the solution of a wide range of computational problems~\cite{nielsen2002quantum}. At the same time, only a small number of quantum algorithmic primitives with provable speedup have been identified, motivating the development of heuristics. Due to the limited availability and imperfections of near-term quantum computers, the design and validation of heuristic quantum algorithms have been largely performed in classical simulation. Additionally, classical simulators are commonly used to validate the results obtained on small-scale near-term devices. As a consequence, fast, high-performance simulators are a crucial tool for algorithm development. 

Quantum Approximate Optimization Algorithm (QAOA)
~\cite{farhi2014quantum,Hogg2000} 
is one of the most promising quantum algorithms for combinatorial optimization. QAOA approximately solves optimization problems by preparing a parameterized quantum state such that upon measuring it, high quality solutions are obtained with high probability. 

Due to the difficulty of theoretical analysis, QAOA performance is commonly analyzed numerically.  Recently, Boulebnane and Montanaro demonstrated numerically that QAOA scales better than state-of-the-art classical solvers for random $8$-SAT~\cite{2208.06909}. The demonstrated potential of QAOA as an algorithmic component that enables quantum speedups motivates the development of tools for its numerical study. Since QAOA performance increases with circuit depth $p$, it is particularly interesting to simulate high-depth QAOA. For example, Ref.~\cite{2208.06909} only observes a quantum speedup with QAOA for $p\gtrsim 14$ and Ref.~\cite{LykovSampling2023} demonstrates that $p\geq12$ is needed for QAOA to be competitive with classical solvers for the MaxCut problem on 3-regular graphs.

\begin{figure}[t]
    \centering
    
    \begin{tikzpicture}[node distance=2em,
        nodes={draw,rounded corners,align=center,
        fill=white,minimum height=3.5em,minimum width=6em}]
        \small
        \tikzstyle{label}=[draw=none,align=center,minimum height=0em,minimum width=0em]
     \node (P)[minimum size=0em, draw=white]{};
     \node[below=4.5em of P] (C) {Precompute\\diagonal};
     \node[right=4em of C] (S) {Calculate\\objective};
     \node[below=4.5em of S] (O) {Optimizer};
      \draw[thick,->] (P) to node[label,pos=0.10] {Cost function to optimize} (C);
        \draw[thick,->] (C) to (S);
        \node[fit=(C) (S), inner sep=8pt, draw=UBCblue, fill opacity=0.1, ] (sim) {};

        \node (simLabel) at (2,-3.25) [label, above=-0.5em of sim] {Simulator};
        \draw[thick,->] (S) to [bend right=40] node[label,pos=0.55] {$\langle \vec\gamma \vec\beta \rvert {\hat C} \lvert \vec\gamma \vec\beta \rangle$} (O);
        \draw[thick,->] (O) to [bend right=40] node[label,pos=0.45] {$\vec\gamma,\vec\beta$} (S);
            \end{tikzpicture}
            
    \caption{Overview of the simulator. Precomputing and storing the diagonal cost operator reduces the cost of both simulating the phase operator in QAOA as well as evaluating the QAOA objective.
    \ifarxiv
    \vspace{-0.2in}
    \fi
    }
    \label{fig:diagram}
\end{figure}
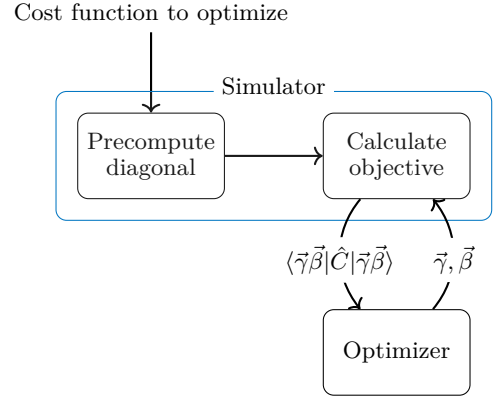

We implement a fast state-vector simulator for the study of QAOA. Our simulator is optimized for simulating QAOA with high depth as well as repeated evaluation of QAOA objective, which is required for tuning the QAOA parameters. To accelerate the simulation, we first precompute the values of the function to be optimized (see Fig.~\ref{fig:diagram}). The result of precomputation is reused during the parameter optimization. The precomputation algorithm is easy to parallelize, making it amenable to GPU acceleration. The precomputation requires storing an exponentially-sized vector, increasing the memory footprint of the simulation by only $12.5\%$. %
Our technique is general, and we implement transverse-field and Hamming-weight-preserving $\xgate\ygate$ mixers. We achieve orders of magnitude speedups over state-of-the-art state-vector and tensor-network simulators and demonstrate scalability to 1,024 GPUs~\cite{2308.02342}. We implement the developed simulator in \qokit{} framework, which also
provides optimized parameters and additional tooling for a set of commonly studied problems 

We use the developed simulator to simulate QAOA with up to $40$ qubits, enabling a scaling analysis of QAOA performance on the LABS problem. The details of the observed quantum speedup over state-of-the-art classical solvers are described in detail in Ref.~\cite{2308.02342}.

\section{Background}
    Consider the problem of minimizing a cost function $f:\mathcal F \to \mathbb R$ defined on a subset $\mathcal{F}$ of the Boolean cube $\mathbb{B}^n$. The bijection $\mathbb{B} \cong \{-1, 1\}$ is used to express the cost function $f$ as a polynomial in terms of spins
    \begin{equation}
        f(\mathbf s) = \sum_{k=1}^{L} w_k\prod_{i\in \mathbf t_k}s_i, \;\;\;\;\; s_i\in\{-1,1\}.
        \label{eq:objective_polynomial}
    \end{equation} 
    \begin{tolerant}{9999}
        The polynomial is defined by a set of terms $\mathcal T = \{(w_1, \mathbf{t_1}), (w_2, \mathbf{t_2}), \dots (w_L, \mathbf{t_L})\}$. Each term consists of a weight $w_k \in \mathbb{R}$ and a set of integers $\mathbf{t_k}$ from $1$ to $n$, i.e., $\mathbf t_k \subseteq \{i\ |\ 1 \leq i\leq n\}$. Constant offset is encoded using a term $(w_{\text{offset}, \emptyset})$.
    \end{tolerant}

    We present numerical results for QAOA applied to the following two problems. First, we consider the commonly studied MaxCut problem. The cost function for the MaxCut problem is given by $\sum_{i,j\in E}\frac{1}{2}s_is_j-\frac{|E|}{2}$
    where $G = (V,E)$ is the problem graph,
    $\mathcal T = E$, and $s_i\in \{-1,1\}$ are the variables to be optimized. Second, we consider the Low Autocorrelation Binary Sequences (LABS) problem. The cost function for the LABS problem with $n$ variables is given by $2\sum_{i=1}^{n-3}s_i \sum_{t=1}^{\lfloor \frac{n-i-1}{2}\rfloor}\sum_{k=t+1}^{n-i-t}s_{i+t}s_{i+k}s_{i+k+t} +\sum_{i=1}^{n-2}s_i\sum_{k=1}^{\lfloor \frac{n-i}{2}\rfloor}s_{i+2k}$.

    The QAOA state is prepared by applying phase and mixing operators in alternation. 
    The phase operator is diagonal and adds phases to computational basis states based on the values of the cost function. The mixing operator, also known as the mixer, is non-diagonal and is used to induce non-trivial dynamics.
     The phase operator is created using the diagonal problem Hamiltonian given by $\hat C = \sum_{
     \mathbf{x}\in\mathcal{F}} f(\mathbf x) \ketbra{\mathbf x}{\mathbf x}$, where $\ket {\mathbf x}$ is a computational basis quantum state. The spectrum of the operator matches the values of the cost function to be optimized, thus the ground state $\ket{\mathbf x^*}$ of such Hamiltonian corresponds to the optimal value $f(\mathbf x^*) $. The goal of QAOA is to bring the quantum system close to a state
     such that upon measuring it, we obtain $\mathbf x^*$ with high probability.
    
    The QAOA circuit is given by 
    $$\ket{\vec\gamma \vec\beta} = \prod_{l=1}^p \left(e^{-\iu\beta_l\hat M} e^{-\iu\gamma_l \hat C}\right) \ket{s}.$$ 
    If $\mathcal{F}=\mathbb{B}^n$, the standard choices are the transverse-field operator $\hat M = \sum_i \xgate_i$ as the mixer and uniform superposition $\ket + ^{\otimes n}$ as the initial state. The free parameters $\gamma_l$ and $\beta_l$ are chosen to minimize the expected solution quality $\langle \vec\gamma \vec\beta \rvert \hat C \lvert \vec\gamma \vec\beta \rangle$, typically using a local optimizer.
    
    \section{Simulation of QAOA using \qokit{}}
    
    The convention of representing a quantum program as a sequence of quantum gates may pose limitations when simulating QAOA classically.
    In standard gate-based simulators such as Qiskit and QTensor, the phase operator must be compiled into gates. The number of these gates typically scales polynomially with the number of terms in the cost function $|\mathcal T|$.
    The overhead is especially large when considering objectives with higher order terms, such as $k$-SAT with $k>3$ and Low Autocorrelations Binary Sequences (LABS) problem. 
    Existing state-vector simulators primarily work by iterating
    over each gate in the circuit and modifying the state vector.
    By exploiting the structure of the circuit, our state-vector
    simulator recognizes that each application of the 
    phase operator involves the same set of gates and that the set acts as a diagonal operator, reducing the cost of simulation.

\subsection{Precomputation of the cost vector}
    \qokit{} precomputes the diagonal elements in the operator $\hat{C}$, which are the values of the cost function $f$ for each assignment of the input. The values are stored as a $2^n$-sized cost vector, which encodes all the information about the problem Hamiltonian.
    \qokit{} provides simple high-level API which supports both cost functions defined as a polynomial on spins (see Listing~\ref{lst:example}), as well as a Python lambda function.
    For precomputation using polynomial terms (Eq.~\ref{eq:objective_polynomial}), we start by allocating an array of zeroes, and iterate over terms in $\mathcal T$, applying a GPU kernel in-parallel for each element of the array.
    The binary representation of an index of a vector element corresponds to qubit values in a basis state. This allows us to calculate the value of the term using bitwise-XOR and ``population count" operations. The kernel calculates the term value and adds it to a single element of the vector in-place. This has the advantage of locality, which is beneficial for GPU parallelization and distributed computing.
    
    To apply the  phase operator with parameter $\gamma_l$, we perform an element-wise product of the state vector and $e^{-\iu\gamma_l \vec C}$, where $\vec C$ is the cost vector and the exponentiation is applied element-wise.
    After simulating the QAOA evolution, we reuse the precomputed $\vec C$ to evaluate the expected solution quality $\langle \vec\gamma \vec\beta \rvert \hat C \lvert \vec\gamma \vec\beta \rangle$ by taking an inner product between $\vec{C}$ and the QAOA state. 

\subsection{Mixing operator}

    Application of the mixing operator is more challenging than that of the phase operator, and accounts for the vast majority of computational cost in our simulation. We briefly discuss the implementation using the example of the transverse-field mixer. Other mixers are implemented similarly. The transverse-field mixer can be decomposed into products of local gates as $U_M = e^{-\iu\beta\sum_i \xgate_i} = \prod_i e^{-\iu\beta \xgate_i}$. Each gate $e^{-\iu\beta \xgate} = \cos(\beta)\idgate - \iu\sin(\beta)\xgate$ ``mixes" two probability amplitudes, and all $n$ gates mix all $2^n$ probability amplitudes. Classical simulation of this operation requires all-to-all communication, where each output vector element depends on every entry of the input vector. For example, for $\beta=\pi/2$ the phase operator implements the Walsh-Hadamard transform, which is a Fourier transform on the Boolean cube $\mathbb B ^n$. 
    The definition of QAOA mixing operator via Walsh-Hadamard transform was known for a long time, see e.g., Refs.~\cite{Hogg2000,hogg2022coloring}. In fact, the ability of quantum computers to efficiently perform Walsh-Hadamard transform~\cite{Brassard1998,HOGG1999}, which is the central building block for the famed Grover's algorithm~\cite{Grover1997}, was the inspiration for the definition of the QAOA mixer~\cite{Hogg2000}.

    Our GPU simulator implements each $e^{-\iu\beta \xgate_i}$ of the mixing operator by applying a GPU kernel which modifies two elements of the state vector. Since these calculations do not interfere with each other, the updates on all pairs of elements in the state vector can be done in place and in parallel, hence well-utilizing the parallelization power of the GPU.
    The algorithm for simulating a single $e^{-\iu\beta \xgate_i}$ is described in Algorithm~\ref{algorithm:inplace-fsu2}. To simulate the full mixer, Algorithm~\ref{algorithm:inplace-fsu2} is applied to each qubit $i \in [n]$, as shown in Algorithm~\ref{algorithm:inplace-fusu2}. Both algorithms modify the state vector in-place without using any additional memory.

    The full QAOA simulation algorithm in \qokit{} is described in Algorithm~\ref{algorithm:fast-qaoa}. Furthermore, we implement the simulation using NVIDIA cuQuantum framework, by replacing Algorithm~\ref{algorithm:inplace-fusu2} with calls to the cuStateVec library. We refer to this implementation as QOKit (cuStateVec).
    In addition to the conventional transverse-field mixing Hamiltonian $M = \sum_i \xgate_i$, we implement Hamming-weight-preserving $\xgate\ygate$ mixer whose Hamiltonian is given by a set of two-qubit operators $M = \sum_{\langle i, j\rangle} \frac{1}{2}(\xgate_i\xgate_j + \ygate_i\ygate_j)$ for $\langle i, j \rangle$ corresponding to the edges of ring or complete graphs. The implementation leverages the observation that Algorithms~\ref{algorithm:inplace-fsu2} and ~\ref{algorithm:inplace-fusu2} can be easily extended to SU(4) operators.

    \begin{algorithm}[h]
        \caption{Fast $\mathrm{SU}(2)$ On A State Vector}
        \label{algorithm:inplace-fsu2}
        \begin{algorithmic}[1]
        \STATEx {\bfseries Input:} Vector $\boldsymbol{x} \in \mathbb{C}^{N}$ with $N = 2^n$, a unitary matrix
        $U_\star =\begin{pmatrix}
            {a} & -{b}^* \\
            {b} & {a}^*
        \end{pmatrix}
        \in \mathrm{SU}(2)$ and a positive integer $d \in [n]$
        \STATEx {\bfseries Output:} Vector $\boldsymbol{y} = U\boldsymbol{x}$, where $U = \idgate^{\otimes (d-1)} \otimes U_\star \otimes \idgate^{\otimes (n-d)}$ and $\idgate$ is the $2$-dimensional identity matrix
    
        \STATE Create a reference $\boldsymbol y$ to input vector $\boldsymbol{x}$
        \FOR{$k_1=1$ {\bfseries to} $2^{n-d}$}
            \FOR{$k_2=1$ {\bfseries to} $2^{d-1}$}
                \STATE Compute indices:
                    \STATEx \hspace{4.0em} $l_1 \leftarrow (k_1-1)2^d+k_2$
                    \STATEx \hspace{4.0em} $l_2 \leftarrow (k_1-1)2^d+k_2+2^{d-1}$
                \STATE Simultaneously update $\boldsymbol{y}_{l_1}$ and $\boldsymbol{y}_{l_2}$:
                    \STATEx \hspace{4.0em} $\boldsymbol{y}_{l_1} \leftarrow {a} \boldsymbol{y}_{l_1} - {b}^* \boldsymbol{y}_{l_2}$
                    \STATEx \hspace{4.0em} $\boldsymbol{y}_{l_2} \leftarrow {b} \boldsymbol{y}_{l_1} + {a}^* \boldsymbol{y}_{l_2}$                
            \ENDFOR
        \ENDFOR
        \STATE {\bfseries return} $\boldsymbol{y}$
        \end{algorithmic}
    \end{algorithm}

    \begin{algorithm}[h]
        \caption{Fast Uniform $\mathrm{SU}(2)$ Transform (Single-node)}
        \label{algorithm:inplace-fusu2}
        \begin{algorithmic}[1]
        \STATEx {\bfseries Input:} Vector $\boldsymbol{x} \in \mathbb{C}^{N}$, a unitary matrix $U \in \mathrm{SU}(N)$ decomposable into tensor product of $n$ unitary matrices in $\mathrm{SU}(2)$, i.e. $U = \bigotimes_{i=1}^{n} U_i = U_n \otimes \dots \otimes U_2 \otimes U_1$, where 
        $U_i =\begin{pmatrix}
            {a}_i & -{b}_i^* \\
            {b}_i & {a}_i^*
        \end{pmatrix}
        \in \mathrm{SU}(2)$ and $N = 2^n$
        \STATEx {\bfseries Output:} Vector $\boldsymbol{y} = U\boldsymbol{x}$
    
        \STATE Create a reference $\boldsymbol y$ to input vector $\boldsymbol{x}$
        \FOR{$i=1$ {\bfseries to} $n$}
            \STATE Apply Algorithm~\ref{algorithm:inplace-fsu2} with $U_\star \leftarrow U_i$ and $d \leftarrow i$
        \ENDFOR
        \STATE {\bfseries return} $\boldsymbol{y}$
        \end{algorithmic}
    \end{algorithm}
    
    \begin{algorithm}[h]
        \caption{Fast Simulation of QAOA}
        \label{algorithm:fast-qaoa}
        \begin{algorithmic}[1]
        \STATEx {\bfseries Input:} Initial vector $\boldsymbol{x} \in \mathbb{C}^{N}$, QAOA circuit parameters $\boldsymbol{\beta}, \boldsymbol{\gamma} \in \mathbb{R}^p$, cost function $f: \mathbb{Z}_2^n \to \mathbb{R}$, where $N = 2^n$
        \STATEx {\bfseries Output:} State vector after applying the QAOA circuit to $\boldsymbol{x}$
    
        \STATE Pre-compute (and cache) cost values for all binary strings into a vector $\boldsymbol{c} \in \mathbb{C}^{N}$
    
        \STATE Initialize output vector $\boldsymbol{y} \leftarrow \boldsymbol{x}$
        
        \FOR{$l=1$ {\bfseries to} $p$}
            \STATE Apply phase operator:
            \STATEx \hspace{2.0em} {\bfseries for} $k=1$ {\bfseries to} $N$ {\bfseries do}
                \STATEx \hspace{3.0em} $y_k \leftarrow e^{-\iu\gamma_l c_k} y_k$
            \STATEx \hspace{2.0em} {\bfseries end for}
            \STATE Apply mixing operator:
            \STATEx \hspace{2.0em} Apply Algorithm~\ref{algorithm:inplace-fusu2} on $\boldsymbol{y}$ with $a_i \leftarrow \cos \beta_l$, $b_i \leftarrow \sin \beta_l$ $\forall\ i \in [n]$
        \ENDFOR
        \STATE {\bfseries return} $\boldsymbol{y}$
        \end{algorithmic}
    \end{algorithm}

\subsection{Distributed simulation}
\label{sec:distributed_methods}
    A typical supercomputer consists of multiple identical compute nodes connected by a fast interconnect. %
    Each node in turn consists of a CPU and several GPUs.
    Since GPUs are much faster in our simulation tasks, we do not use CPUs in our distributed simulation. Each of $K$ GPUs holds a slice of the state vector, which corresponds to fixing the values of a set of $k = \log_2(K)$ qubits. For example, for $K=2$ GPUs, the first GPU holds probability amplitudes for states with the first qubit in state $\ket 0$, while the second GPU holds states with first qubit in the state $\ket 1$. In general, using $K$ GPUs allows us to increase the simulation size by $k$ qubits.
    
    During the precomputation, the cost vector $\vec C$ is sliced in the same  way as the state vector. Due to the locality discussed above, the precomputation and the phase operator application  do not  require any communication across GPUs. The most expensive part of the  simulation is the mixing operator, since it requires an all-to-all communication pattern.
    In our simulation we distribute the state vector by splitting it into $K$ chunks, which corresponds to fixing first $k$ qubits, which we call \emph{global qubits}. Bits of the binary representation of the node index determine the fixed qubit values. The remaining $n- k$ qubits are referred to as \emph{local qubits}.
    
    The mixer application starts by applying the $e^{-\iu\beta \xgate_i}$ gates that correspond to local qubits.
    To apply $\xgate$ rotations on global qubits, we reshape the distributed state vector using the \texttt{MPI\_Alltoall} MPI collective.
    This operation splits each local state vector further into $K$ subchunks and transfers subchunk $A$ of process $B$ into subchunk $B$ of process $A$.
    If each subchunk consists of one element and we arrange the full state vector in a matrix with process id as column index and subchunk id as the row index, then the call to \texttt{MPI\_Alltoall} performs a transposition  of this matrix. For a $n$-qubit simulation the algorithm requires $2k \leq n$ 
    to ensure that there is at least one element in each subchunk. Consider the state vector reshaped as a tensor $V_{abc}$ with $a$ being the process id representing the $k$ global qubits, $b$ being a multi-index of first $k$ local qubits, and $c$ being a multi-index of the last $n - 2 k$ qubits. Then the \texttt{MPI\_Alltoall} operation corresponds to a transposition of the first two indices, i.e., $V_{abc} \to V_{bac}$. Thus, after this transposition, the global qubits become local and we are free to apply operations on those $k$ global qubits locally in each process. The algorithm concludes by applying the \texttt{MPI\_Alltoall} once again to restore  the original qubit ordering.
    This algorithm is described in Algorithm~\ref{algorithm:inplace-fusu2-parallel}.

     \begin{algorithm}[h]
        \caption{Fast Uniform $\mathrm{SU}(2)$ Transform (Multi-node)}
        \label{algorithm:inplace-fusu2-parallel}
        \begin{algorithmic}[1]
        \STATEx {\bfseries Input:} Vector $\boldsymbol{x} \in \mathbb{C}^{N}$ distributed over $K$ nodes, a unitary matrix $U = \bigotimes_{i=1}^{n} U_i = U_n \otimes \dots \otimes U_2 \otimes U_1$, where 
        $U_i \in \mathrm{SU}(2)$ and $N = 2^n$
        \STATEx {\bfseries Output:} Distributed vector $\boldsymbol{y} = U\boldsymbol{x}$
    
        \STATE Create a reference $\boldsymbol y$ to the local slice of input vector $\boldsymbol{x}$
        \FOR{$i=1$ {\bfseries to} $n - \log_2 K$}
            \STATE Apply Algorithm~\ref{algorithm:inplace-fsu2} with $U_\star \leftarrow U_i$ and $d \leftarrow i$ to the local slice $\boldsymbol y$.
        \ENDFOR
        \STATE Run in-place \texttt{MPI\_AlltoAll} on the local slice $\boldsymbol y$.
        \FOR{$i=n-\log_2 K +1$ {\bfseries to} $n$}
            \STATE Apply Algorithm~\ref{algorithm:inplace-fsu2} with $U_\star \leftarrow U_i$ and $d \leftarrow i-\log_2 K$ to the local slice $\boldsymbol y$.
        \ENDFOR
        \STATE Run in-place \texttt{MPI\_AlltoAll} on the local slice $\boldsymbol y$.
        \STATE {\bfseries return} $\boldsymbol{y}$
        \end{algorithmic}
    \end{algorithm}

    The \texttt{MPI\_Alltoall} is known be a challenging collective communication routine, since it requires the total transfer of the full state vector $K$ times. There exist many algorithms for this implementation~\cite{visual_mpi, mpi_collective_performance}, each with its own trade-offs. Furthermore, the same communication problem occurs in applying distributed Fast Fourier Transform (FFT), which has been studied extensively~\cite{heffte, gholami2016accfft, multi_process_fft,ayala2021interim}. In this work, we use the out-of-the-box MPI implementation Cray MPICH. Utilizing the research on distributed FFT may help further improve our implementation.

\section{Examples of use}
    
    \qokit{} consists of two conceptual parts:
    \begin{enumerate}
        \begin{tolerant}{9999}
        \item Low-level simulation API defined by an abstract class
        \texttt{qokit.fur.QAOAFastSimulatorBase}
        \end{tolerant}
        \item Easy-to-use one-line methods for simulating MaxCut, LABS and portfolio optimization problems
    \end{enumerate}

    \begin{tolerant}{9999}
    The low-level simulation API is designed to provide more flexibility in terms of inputs, methods and outputs of simulation. The simulation inputs can be specified by providing either terms $\mathcal T$
    or existing pre-computed diagonal vector. The simulation method is specified by using a particular subclass of
    \texttt{qokit.fur.QAOAFastSimulatorBase}
    or by using a shorthand method
    \texttt{qokit.fur.choose\_simulator}.
    This simulator class is the main means of simulation, with input parameters being passed in the constructor, the simulation done in \texttt{simulate\_qaoa} method, and outputs type specified by choosing a corresponding method of the simulator object. An example of using the simulator with input \texttt{terms} parameter is shown in Listing~\ref{lst:example}.
    \end{tolerant}
    \vspace{1em}
    {
    \centering
    \begin{minipage}{\linewidth}
    \lstset{basicstyle=\linespread{1.3}\footnotesize\ttfamily,breaklines=true,caption={Evaluating the QAOA objective for weighted MaxCut problem on an all-to-all graph using \qokit{}.},label=lst:example} 
    \begin{lstlisting}
  import qokit
  simclass = qokit.fur.choose_simulator(name='auto')
  n = 28  # number of qubits
  # terms for all-to-all MaxCut with weight 0.3
  terms = [(.3, (i, j)) for i in range(n) for j in range(i+1, n)]
  sim = simclass(n, terms=terms)
  # get precomputed cost vector 
  costs = sim.get_cost_diagonal()
  result = sim.simulate_qaoa(gamma, beta)
  E = sim.get_expectation(result)
        \end{lstlisting}
    \end{minipage}
    }

    \qokit{} implements five different simulator classes that share the same API:
    \begin{enumerate}
        \item \texttt{python} -- A portable CPU \texttt{numpy}-based version
        \item \texttt{c} -- Custom CPU simulator implemented in C

        \item \texttt{nbcuda} -- GPU simulator using \texttt{numba}
        \item \texttt{gpumpi} -- A distributed version of the GPU simulator
        \item \texttt{cusvmpi} -- A distributed GPU simulator with cuStateVec as backend
    \end{enumerate}

    To choose from the simulators, one may use one of the following three methods, depending on the choice of mixer type:
    \begin{enumerate}
        \item \texttt{qokit.fur.choose\_simulator()}
        \item \texttt{qokit.fur.choose\_simulator\_xyring()}
        \item \texttt{qokit.fur.choose\_simulator\_xycomplete()}
    \end{enumerate}
    Each of these simulators accepts an optional \texttt{name} parameter. The default simulator is chosen based on existence of GPU or configured MPI environment.
    An example of using a custom mixer for simulation is provided in Listing~\ref{lst:custom_mixer}.
    
    \vspace{1em}
    {
    \centering
    \begin{minipage}{\linewidth}
    \lstset{basicstyle=\linespread{1.3}\footnotesize\ttfamily,breaklines=true,
    caption={Using \qokit{} with a different mixing operator: $M=\sum_{\langle i, j\rangle} \frac{1}{2}(\xgate_i\xgate_j + \ygate_i\ygate_j)$ for tuples $\langle i,j \rangle$ from a complete graph on qubits.}
    ,label=lst:custom_mixer}
        \begin{lstlisting}
  import qokit
  simclass = qokit.fur.choose_simulator_xycomplete()
  n = 40
  terms = qokit.labs.get_terms(n)
  sim = simclass(n, terms=terms)
  result = sim.simulate_qaoa(gamma, beta)
  E = sim.get_expectation(result)
        \end{lstlisting}
    \end{minipage}
    }
    
    The constructor of each simulator class accepts one of \texttt{terms} or \texttt{costs} argument. The \texttt{terms} argument is a list of tuples $(w_k, \mathbf t_k)$, where $w_k$ is the weight of product defined by $\mathbf t_k$, whih is a tuple of integers specifying the indices of Boolean variables involved in this product, as described in Equation~\ref{eq:objective_polynomial}. 
    The simulation method returns a \texttt{result} object, which is a representation of the evolved state vector. The data type of this object may change depending on simulator type, and for best portability it is advised to use the output methods instead of directly interacting with this object.
    The output methods all have \texttt{get\_} prefix, accept the \texttt{result} object as their first argument, and return CPU values. These methods are:
    
    \begin{enumerate}
        \item \texttt{get\_expectation(result)}
        \item \texttt{get\_overlap(result)}
        \item \texttt{get\_statevector(result)}
        \item \texttt{get\_probabilities(result)}
    \end{enumerate}
    
    When evaluating the expectation and overlap with the ground state, the cost vector from the phase operator is used by default. This vector is precomputed at the class instantiation and can be retrieved using \texttt{get\_cost\_diagonal()} method. Alternatively, the user may specify a custom cost vector by passing it as the \texttt{costs} argument when calling \texttt{get\_expectation} or \texttt{get\_overlap}.
    
    The output methods may accept additional optional arguments depending on the type of the simulator.
    For example, GPU simulators' \texttt{get\_probabilities} method has \texttt{preserve\_state} argument (default \texttt{True}) which specifies whether to preserve the statevector for additional calculations; otherwise, the norm-square operation will be applied in-place. In both cases, the method returns a real-valued array of probabilities.
    Distributed GPU simulators accept \texttt{mpi\_gather} argument (default \texttt{True}) that signals the method to return a full state vector on each node. Specifying \texttt{mpi\_gather = True} guarantees that the same code will produce the same result if the hardware-specific simulator class is changed. An example of using \qokit{} for distributed simulation is provided in Listing~\ref{lst:example_distributed}.

    \vspace{1em}
    {
    \centering
    \begin{minipage}{\linewidth}
    \lstset{basicstyle=\linespread{1.3}\footnotesize\ttfamily,breaklines=true,
    caption={Evaluating the QAOA objective for LABS problem using MPI on a distributed computing system using \qokit{}. The \texttt{preserve\_state} argument is used to reduce memory usage when evaluating the expectation value.}
    ,label=lst:example_distributed}
        \begin{lstlisting}
  import qokit
  simclass = qokit.fur.choose_simulator(name='cusvmpi')
  n = 40
  terms = qokit.labs.get_terms(n)
  sim = simclass(n, terms=terms)
  result = sim.simulate_qaoa(gamma, beta)
  E = sim.get_expectation(result, preserve_state=False)
        \end{lstlisting}
    \end{minipage}
    }

\section{Performance of \qokit{}}

    We now present a comparison of \qokit{} performance to state-of-the-art state-vector and tensor-network quantum simulators. We show that our framework %
    has lower runtimes and scales well to large supercomputing systems. All reported benchmarks are executed on the Polaris supercomputer accessed through the Argonne Leadership Computing Facility. Single-node results are obtained using a compute node with two AMD EPYC 7713 64-Core CPUs with 2 threads per core, $503$~GB of RAM and an NVIDIA A100 GPU with $80$~GB of memory. In all experiments the state vector is stored with double precision (\texttt{complex128} data type).
    
\subsection{CPU and GPU simulation}\label{sec:results_single_node}
    The CPU simulation is implemented in two ways: using the NumPy Python library and using a custom C code (``\texttt{c}'' simulator above). The latter is more performance, so we only report the results with \texttt{c} simulator.
    We evaluate the CPU performance by simulating QAOA with $p=6$ on MaxCut random regular graphs. 

        \begin{figure}[t]
            \centering
            \includegraphics[width=\linewidth]{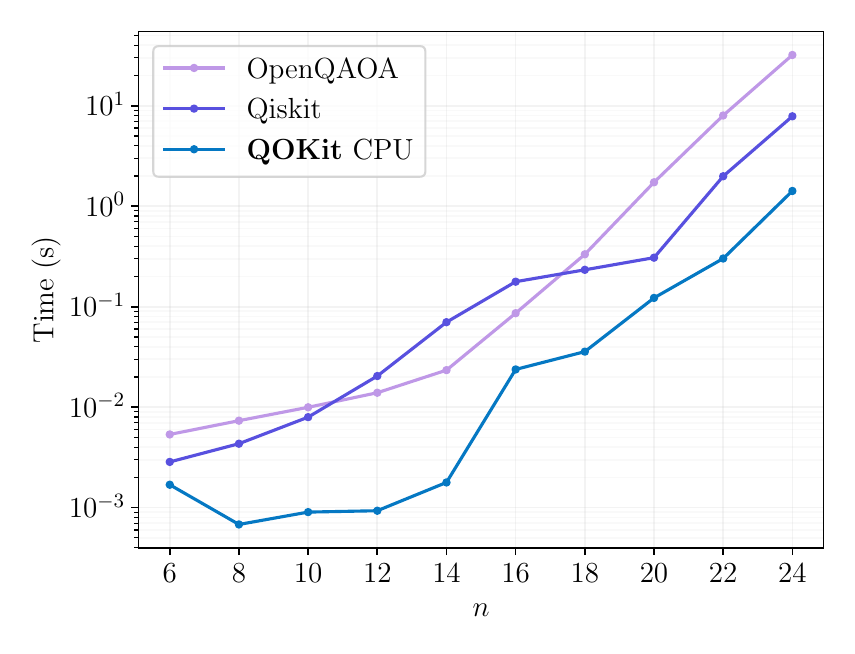}
            \caption{Runtime of end-to-end simulation of QAOA expectation value with $p=6$ on MaxCut problem on 3-regular graphs with commonly-used CPU simulators for QAOA.  They time plotted is the mean over 5 runs.}
            \label{fig:cpu_perf_compare}
        \end{figure}
    
    \begin{figure}[t]
        \centering
        \includegraphics[width=1.\linewidth]{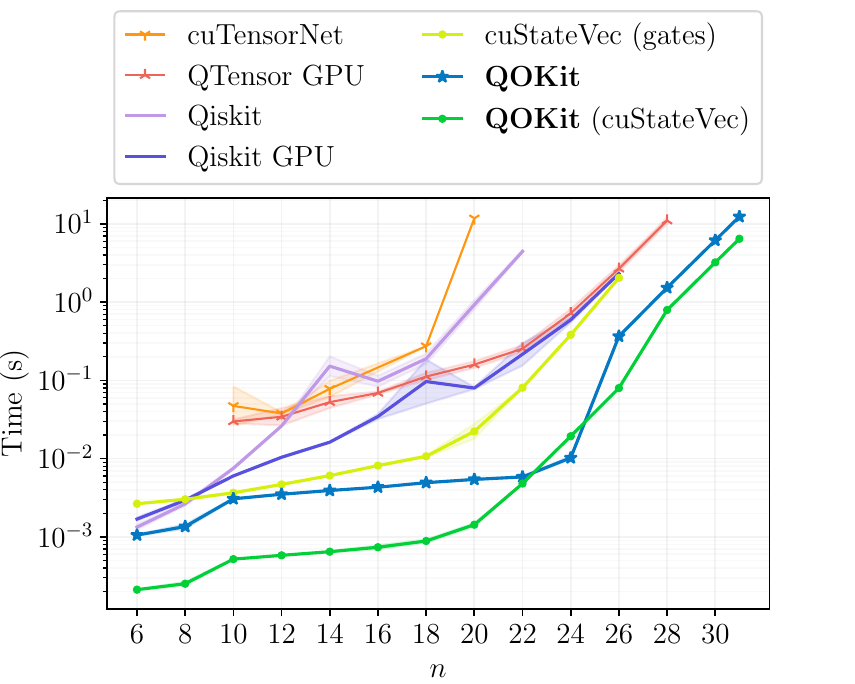}
        \caption{Time to apply a single layer of QAOA for the LABS problem with commonly-used CPU and GPU simulators. \qokit{} simulator uses the precomputation which is not included in current plot. The precomputation time is amortized, as shown on Figure~\ref{fig:amortization}.
        \qokit{} can be configured to use cuStateVec for application of the mixing operator, which provides the best results.
        }
        \label{fig:compare}
    \end{figure}

    Figure~\ref{fig:cpu_perf_compare} shows a comparison of runtime for varying number of qubits for commonly-used CPU simulators. We use \qokit{} \texttt{c} simulator, Qiskit Aer state-vector simulator version 0.12.2, and OpenQAOA ``vectorized'' simulator version 0.1.3. We observe $\approx 5-10\times$ speedup against Qiskit~\cite{Qiskit} and OpenQAOA~\cite{sharma2022openqaoa} across a wide range of values of $n$.
    We note that the simulation method in QAOAKit~\cite{Shaydulin2021QAOAKit} is Qiskit, which is why we do not benchmark it separately.

        \begin{figure}
            \centering
            \vspace{-0.1in} %
            \includegraphics[width=\linewidth]{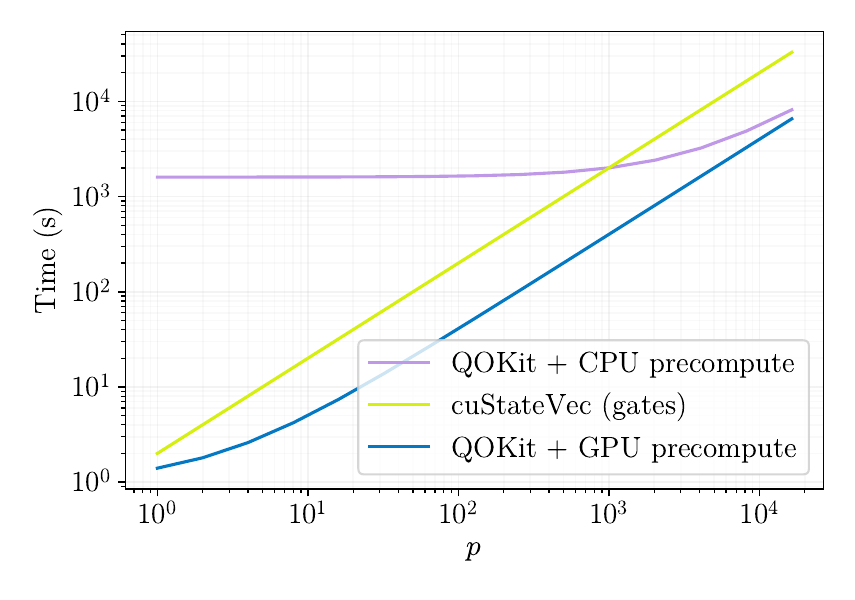}
            \caption{Total simulation time vs. number of layers in QAOA circuit for LABS problem with $n=26$. The GPU precomputation is fast enough to provide speedup over the gate-based state-vector simulation (cuStateVec) even for a single evaluation of the QAOA circuit.}
            \label{fig:amortization}
        \end{figure}

    We evaluate the GPU performance by evaluating time to simulate one layer of QAOA applied to the LABS problem. Fig.~\ref{fig:compare} provides a comparison between \qokit{} and commonly used state-vector (Qiskit~\cite{Qiskit} version 0.43.3, cuStateVec~\cite{cuquantum}) and tensor-network (cuTensorNet~\cite{cuquantum}, QTensor~\cite{QTensor}) simulators. We used CuQuantum Python package version 23.6.0 and cudatoolkit version 14.4.4. 
    The tensor network timing is obtained by running calculation of a single probability amplitude for various values of $1\leq p\leq 15$ and dividing the total contraction time by $p$. Deep circuits have optimal contraction order that produces contraction width equal to $n$. Since obtaining batches of amplitudes does not produce high overhead~\cite{Schutski_adaptive}, this serves as a lower bound for full state evolution. Note that the so-called ``lightcone approach'', wherein only the reverse causal cone of the desired observable is simulated, does not significantly reduce the resource requirements due to the high depth and connectivity of the phase operator.
    For QTensor, ``tamaki\_30'' contraction optimization algorithm is used. CuTensorNet contraction is optimized with default settings. It is possible that the performance can be improved by using diagonal gates~\cite{Lykov_diagonal_gates}, which are only partially supported by cuTensorNet at this moment.
    
    For $n>20$, we observe that the precomputation provides orders of magnitude speedups for simulation of a QAOA layer. The LABS problem has a large number of terms in the cost function, leading to deep circuits which put tensor network simulators at a disadvantage. As a consequence, we observe that tensor network simulators are slower than state-vector simulation. We also observe that using cuStateVec as a backend for mixer gate simulation provides additional $\approx 2\times$ speedup, possibly due to higher numerical efficiency achieved by in-house NVIDIA implementation. We do not include the precomputation cost in Fig.~\ref{fig:compare}. This cost is amortized over application of a QAOA layer as shown in Fig.~\ref{fig:amortization}, and is negligible if precomputation is performed on GPU. Simulation of each layer in a deep quantum circuit has the same time and memory cost. Thus, to obtain the time for multiple function evaluations, one can simply use this plot with aggregate number of layers in all function evaluations.

    Our best GPU performance for QAOA on LABS problem is $\approx 6$ seconds per QAOA layer for $n=31$ using double precision. This simulation requires the same memory amount as one with $n=32$ using single precision.
    In addition to our own implementation, we benchmark the same simulator as in the Ref.~\cite{bayraktar2023cuquantum} on the LABS problem. For smaller $n\leq 26$, \qokit{} with cuStateVec shows a $\approx20\times$ speedup from our precomputation approach. We discuss the choice of cuStateVec as the baseline as well as other state-of-the-art simulation techniques in Sec.~\ref{sec:related}.

        \begin{figure}[t]
            \centering
            \includegraphics[width=\linewidth]{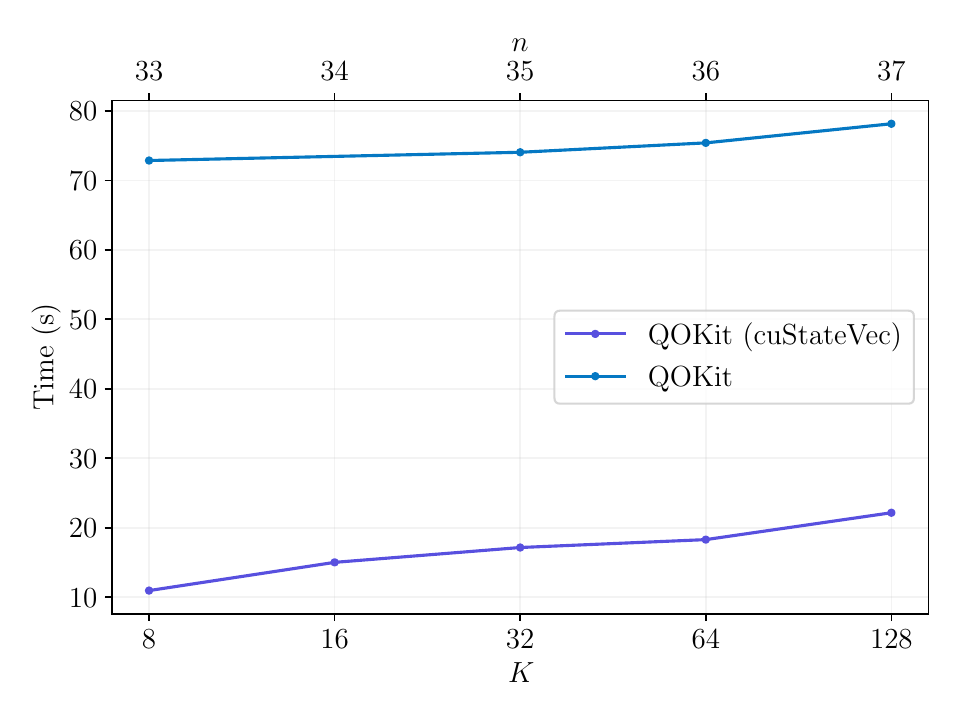}
            \caption{Weak scaling results for simulation of 1 layer of LABS QAOA on Polaris supercomputer. We observe that cuStateVec backend has lower communication overhead, leading to lower overall runtime.}
            \label{fig:distributed_results}
        \end{figure}
        
\subsection{Distributed simulation}

    Finally, we scale the QAOA simulation to $n=40$ qubits using 1024 GPUs of the Polaris supercomputer. %
    At $n=40$, we observe a runtime of $\approx 20$~s per layer. The results of the simulation are discussed in detail in Ref.~\cite{2308.02342}. Here, we focus on the technical aspects of the simulation.

    In distributed experiments, we use compute nodes of Polaris with 4 NVIDIA A100 GPUs with 40GB of memory. 
    The maximum values of $f$ are known for $n<65$, and they are less then $2^{16}$. Therefore, we are able to store the precomputed diagonal as a $2^n$ vector of $\texttt{uint16}$ values, which reduces the memory overhead of the cost value vector.
    As discussed in Section~\ref{sec:distributed_methods}, the most expensive part of this simulation is communication. We implement two approaches for this simulation, a custom MPI code that uses \texttt{MPI\_Alltoall} collective and an implementation leveraging the distributed index swap operation in cuStateVec. Weak scaling results in Figure~\ref{fig:distributed_results} demonstrate the advantage of cuStateVec implementation of communication. The GPUs co-located on a single node are connected with high-bandwidh NVLink network. To transfer data between nodes, GPU data need to transfer to CPU for subsequent transfer to another node. This requires the communication to correctly choose the communication method depending on GPU location. MPI has built-in support which can be enabled using \texttt{MPI\_GPU\_SUPPORT\_ENABLED} environment variable. However, it shows worse performance than the cuStateVec communication code, which uses direct CUDA peer-to-peer communication calls for local GPU communication. We observe that our performance is comparable to distributed simulation reported in Ref.~\cite{bayraktar2023cuquantum}, despite having $2\times$ fewer GPUs per node. This is due to the majority of time being spent in communication, which is confirmed by our smaller-scale profiling experiments. Further research, including adapting the communication patterns used in high-efficiency FFT algorithms, may improve our results.

\section{Related work}\label{sec:related}

    Classical simulation of quantum systems is a dynamic field with a plethora of simulation algorithms~\cite{Dahi2023ASO, QTensor, Gray_2021, zulehner2019advanced_decision_diagrams} and a variety of use cases~\cite{burgholzer2020advanced, lykov_step_dependent_qtensor}. The main approaches to simulation are tensor network contraction algorithms and state-vector evolution algorithms.
    Tensor network algorithms are able to utilize the structure of quantum circuit and need not store the full $2^n$-dimensional state vector when simulating $n$ qubits. Instead, they construct a tensor network and contract it in the most efficient way possible. This approach works best when the circuit is shallow, since the tensor network contraction can be performed across qubit dimension instead of over time dimension. The main research areas of this approach are finding the best contraction order~\cite{cameron_contraction_trees, khakhulin2020learning, Meirom2022OptimizingTN} and applying approximate simulation algorithms~\cite{gray2022hyperoptimized, Contracting_arbitrary_tensor_networks}. However, while there is no theoretical limitation on simulating deep circuits using tensor networks, it is challenging to implement a performant simulator of deep quantum circuits based on tensor networks, as demonstrated by the numerical experiments above.
    
    The state vector evolution algorithms are more straightforward and intuitive to implement. The main limitation of state-vector simulator is the $2^n$ size of the state vector. There are many approaches to improve state-vector simulators. Compressing the state vector has been proposed to reduce the memory requirement and enable the simulation of a higher number of qubits~\cite{full_state_compression}.
    To utilize the structure in the set of quantum gates, some state-vector simulators use the gate fusion approach~\cite{cirq_approximate, tensorflow_quantum, Efthymiou2022, Smelyanskiy2016qHiPSTERTQ}.
    The idea is to group the gates that act on a set of $F$ qubits,
    then create a single $F$-qubit gate by multiplying these gates together. This approach is often applied for $F=2$ and provides significant speed improvements. Computing the fused gate requires storing $4^{F}$ complex numbers, which is a key limitation. Our approach corresponds to using gate fusion with $F=n$, but the group of gates is known to produce a diagonal gate, which can be stored as a vector of only $2^n$ elements.
    
    In our comparison in Sec.~\ref{sec:results_single_node}, we do not enable gate fusion in cuStateVec. While gate fusion may improve the performance of cuStateVec, we believe it is 
    very unlikely to achieve the same efficiency as our method of using the precomputed diagonal cost operator. Our argument is based on examining the results reported in Ref.~\cite{bayraktar2023cuquantum}.
    The central challenge for gate fusion is presented by the fact that the phase operator for the LABS problem requires many gates to implement.
    For example, for $n=31$, the LABS cost function has $\approx 75n$ terms, with many of them being 4-order terms.
    If decomposed into 2-qubit gates, the circuit for QAOA with $p=1$ for the LABS problem has $\approx 160n$ gates after compilation. For comparison, QAOA circuit from Ref.~\cite{bayraktar2023cuquantum} for $n=33$ and $p=2$ has $50n$ un-fused gates, which is $\approx 7\times$ fewer.
    After gate fusion, the circuits from Ref.~\cite{bayraktar2023cuquantum} have $\approx 4n$ fused gates.
    Our precomputation approach reduces the number of gates to practically only the $n$ mixer gates. Thus, assuming that application of a single gate takes the same amount of time for any statevector simulator, we can expect a speedup in the range of $4-160\times$. We note that while this estimate 
    does not take into account various other time factors and variation in gate application times, it provides intuition for why we rule out gate fusion outperforming our techniques.
    
    While there exist a multitude of quantum simulation frameworks, the best results for state-vector GPU simulation that we found in the literature were reported in Ref.~\cite{bayraktar2023cuquantum} (cuQuantum) and Ref.~\cite{cirq_approximate} (qsim). Ref.~\cite{bayraktar2023cuquantum} reports $\approx 10$ seconds for simulating QAOA with $p=2$, $n=33$ qubits and $1650$ gates, using the \texttt{complex64} data type on a single A100 GPU with $80$~GB memory. Ref.~\cite{cirq_approximate} presents single-precision simulation results on a A100 GPU with $40$~GB memory. The benchmark uses random circuits 
    of depth of $20$. The reported simulation time for $n=32$ is $\approx 6$~seconds. Notably, this time may depend significantly on the structure of the circuit since it impacts the gate fusion, as shown in Ref.~\cite{tensorflow_quantum}. Assuming similar gate count, these results show very similar performance, since the simulation in Ref.~\cite{bayraktar2023cuquantum} is two times larger. This motivates our use of cuQuantum (Ref.~\cite{bayraktar2023cuquantum}) as a baseline state-of-the-art state-vector quantum simulator.

 Symmetry of the function to be optimized has been shown to enable a  reduction in the computational and memory cost of QAOA simulation~\cite{Shaydulin2021,Shaydulin2021TQE,2211.09270}. While we do not implement symmetry-based optimizations in this work, they can be combined with our techniques to further improve performance.

    In addition to the simulation method, many simulators differ in the scope of the project. Some simulators like cuQuantum~\cite{bayraktar2023cuquantum} position themselves as a simulator-development SDK, with flexible but complicated API. On the other hand, there exist simulation libraries that focus on the quantum side and delegate the concern of low-level performance to other libraries~\cite{sharma2022openqaoa, Shaydulin2021QAOAKit}. Many software packages exist somewhere in the middle, featuring full support for quantum circuit simulation and focusing on end-to-end optimization efforts on a particular quantum algorithm or circuit type~\cite{lykov_step_dependent_qtensor, gray2022hyperoptimized}. \qokit{} is positioned as one of such packages, as it provides both an optimized low-level QAOA-specific simulation algorithm as well as high-level quantum optimization API for specific optimization problems.

\section{Conclusion}

    We develop a fast and easy-to-use simulation framework for quantum optimization. 
    We apply a simple but powerful optimization by precomputing the values of the cost function. We use the precomputed values to apply the QAOA phase operator by a single elementwise multiplication and to compute the QAOA objective by a single inner product. We provide an easy-to-use high-level API for a range of commonly considered problems, as well as low-level API for extending our code to other problems. We demonstrate orders of magnitude gains in performance compared to commonly used quantum simulators. By scaling our simulator to 1,024 GPUs and 40 qubits, we enabled an analysis of QAOA on the Low Autocorrelation Binary Sequences problem that demonstrated a quantum speedup over state-of-the-art classical solvers~\cite{2308.02342}.

    After this manuscript appeared on arXiv, we became aware of a serial CPU-only Python implementation of a QAOA simulator that uses diagonal Hamiltonian precomputation and Fast Walsh-Hadamard transform to accelerate QAOA state simulation and QAOA objective evaluation~\cite{TQA_github,Sack2021quantumannealing}. We note that Ref.~\cite{TQA_github} requires two applications of fast Walsh-Hadamard transform (forward and inverse) and a diagonal Hamiltonian operation to simulate one layer of QAOA mixer, whereas Algorithms~\ref{algorithm:inplace-fsu2},~\ref{algorithm:inplace-fusu2} apply the mixer in one step with a cost equivalent to one application of fast Walsh-Hadamard transform.
    In addition, the implementation of fast Walsh-Hadamard transform in Ref.~\cite{TQA_github} requires one additional copy of the input state vector, whereas Algorithms~\ref{algorithm:inplace-fsu2},~\ref{algorithm:inplace-fusu2} applies the mixer in place.

\section{Acknowledgements}

The authors thank Tianyi Hao, Zichang He and Minzhao Liu for their invaluable feedback, technical advice and contributions to the \texttt{QOKit} package. The authors thank Alexander Buts and Terence Moore for their support in packaging and releasing \texttt{QOKit}. The authors thank their colleagues at the Global Technology Applied Research center of JPMorgan Chase for support and helpful discussions. The authors thank Yiren Lu for pointing out Ref.~\cite{TQA_github} to the authors. This work used in part the resources of the Argonne Leadership Computing Facility, which is a Department of Energy Office of Science User Facility supported under Contract DE-AC02-06CH11357. The views, opinions and/or findings expressed are those of the authors and should not be interpreted as representing the official views or policies of the Department of Energy or the U.S. Government.

\bibliographystyle{myIEEEtran}
\bibliography{bibliography}

\begin{thebibliography}{10}
\providecommand{\url}[1]{#1}
\csname url@samestyle\endcsname
\providecommand{\newblock}{\relax}
\providecommand{\bibinfo}[2]{#2}
\providecommand{\BIBentrySTDinterwordspacing}{\spaceskip=0pt\relax}
\providecommand{\BIBentryALTinterwordstretchfactor}{4}
\providecommand{\BIBentryALTinterwordspacing}{\spaceskip=\fontdimen2\font plus
\BIBentryALTinterwordstretchfactor\fontdimen3\font minus
  \fontdimen4\font\relax}
\providecommand{\BIBforeignlanguage}[2]{{%
\expandafter\ifx\csname l@#1\endcsname\relax
\typeout{** WARNING: IEEEtran.bst: No hyphenation pattern has been}%
\typeout{** loaded for the language `#1'. Using the pattern for}%
\typeout{** the default language instead.}%
\else
\language=\csname l@#1\endcsname
\fi
#2}}
\providecommand{\BIBdecl}{\relax}
\BIBdecl

\bibitem{nielsen2002quantum}
M.~A. Nielsen and I.~Chuang, \emph{Quantum computation and quantum
  information}, 2002.

\bibitem{farhi2014quantum}
\BIBentryALTinterwordspacing
E.~Farhi, J.~Goldstone, and S.~Gutmann, ``A quantum approximate optimization
  algorithm,'' \emph{arXiv:1411.4028}, 2014. [Online]. Available:
  \href{https://arxiv.org/abs/1411.4028}{1411.4028}
\BIBentrySTDinterwordspacing

\bibitem{Hogg2000}
\BIBentryALTinterwordspacing
T.~Hogg, ``Quantum search heuristics,'' \emph{Physical Review A}, vol.~61,
  no.~5, Apr. 2000.  10.1103/physreva.61.052311. [Online]. Available:
  \url{https://doi.org/10.1103/physreva.61.052311}
\BIBentrySTDinterwordspacing

\bibitem{2208.06909}
\BIBentryALTinterwordspacing
S.~Boulebnane and A.~Montanaro, ``Solving {Boolean} satisfiability problems
  with the quantum approximate optimization algorithm,''
  \emph{arXiv:2208.06909}, 2022. [Online]. Available:
  \href{https://arxiv.org/abs/2208.06909}{2208.06909}
\BIBentrySTDinterwordspacing

\bibitem{LykovSampling2023}
\BIBentryALTinterwordspacing
D.~Lykov, J.~Wurtz, C.~Poole, M.~Saffman, T.~Noel, and Y.~Alexeev, ``Sampling
  frequency thresholds for the quantum advantage of the quantum approximate
  optimization algorithm,'' \emph{npj Quantum Information}, vol.~9, no.~1, Jul.
  2023.  10.1038/s41534-023-00718-4. [Online]. Available:
  \url{https://doi.org/10.1038/s41534-023-00718-4}
\BIBentrySTDinterwordspacing

\bibitem{2308.02342}
\BIBentryALTinterwordspacing
R.~Shaydulin, C.~Li, S.~Chakrabarti, M.~DeCross, D.~Herman, N.~Kumar,
  J.~Larson, D.~Lykov, P.~Minssen, Y.~Sun, Y.~Alexeev, J.~M. Dreiling, J.~P.
  Gaebler, T.~M. Gatterman, J.~A. Gerber, K.~Gilmore, D.~Gresh, N.~Hewitt,
  C.~V. Horst, S.~Hu, J.~Johansen, M.~Matheny, T.~Mengle, M.~Mills, S.~A.
  Moses, B.~Neyenhuis, P.~Siegfried, R.~Yalovetzky, and M.~Pistoia, ``Evidence
  of scaling advantage for the quantum approximate optimization algorithm on a
  classically intractable problem,'' \emph{arXiv:2308.02342}, 2023. [Online].
  Available: \href{https://arxiv.org/abs/2308.02342}{2308.02342}
\BIBentrySTDinterwordspacing

\bibitem{hogg2022coloring}
A.~Fabrikant and T.~Hogg, ``Graph coloring with quantum heuristics,'' in
  \emph{Proceedings of the Eighteenth National Conference on Artificial
  Intelligence and Fourteenth Conference on Innovative Applications of
  Artificial Intelligence}, R.~Dechter, M.~J. Kearns, and R.~S. Sutton,
  Eds.\hskip 1em plus 0.5em minus 0.4em\relax {AAAI} Press / The {MIT} Press,
  2002, pp. 22--27.

\bibitem{Brassard1998}
\BIBentryALTinterwordspacing
G.~Brassard, P.~H{\O}yer, and A.~Tapp, ``Quantum counting,'' in \emph{Automata,
  Languages and Programming}.\hskip 1em plus 0.5em minus 0.4em\relax Springer
  Berlin Heidelberg, 1998, pp. 820--831. [Online]. Available:
  \url{https://doi.org/10.1007/bfb0055105}
\BIBentrySTDinterwordspacing

\bibitem{HOGG1999}
\BIBentryALTinterwordspacing
T.~Hogg, C.~Mochon, W.~Polak, and E.~Rieffel, ``Tools for quantum algorithms,''
  \emph{International Journal of Modern Physics C}, vol.~10, no.~07, pp.
  1347--1361, Oct. 1999.  10.1142/s0129183199001108. [Online]. Available:
  \url{https://doi.org/10.1142/s0129183199001108}
\BIBentrySTDinterwordspacing

\bibitem{Grover1997}
\BIBentryALTinterwordspacing
L.~K. Grover, ``Quantum mechanics helps in searching for a needle in a
  haystack,'' \emph{Physical Review Letters}, vol.~79, no.~2, pp. 325--328,
  Jul. 1997.  10.1103/physrevlett.79.325. [Online]. Available:
  \url{https://doi.org/10.1103/physrevlett.79.325}
\BIBentrySTDinterwordspacing

\bibitem{visual_mpi}
\BIBentryALTinterwordspacing
N.~Netterville, K.~Fan, S.~Kumar, and T.~Gilray, ``A visual guide to {MPI}
  all-to-all,'' in \emph{2022 {IEEE} 29th International Conference on High
  Performance Computing, Data and Analytics Workshop ({HiPCW})}.\hskip 1em plus
  0.5em minus 0.4em\relax {IEEE}, Dec. 2022.  10.1109/hipcw57629.2022.00008.
  [Online]. Available: \url{https://doi.org/10.1109/hipcw57629.2022.00008}
\BIBentrySTDinterwordspacing

\bibitem{mpi_collective_performance}
\BIBentryALTinterwordspacing
J.~Pjesivac-Grbovic, T.~Angskun, G.~Bosilca, G.~Fagg, E.~Gabriel, and
  J.~Dongarra, ``Performance analysis of {MPI} collective operations,'' in
  \emph{19th {IEEE} International Parallel and Distributed Processing
  Symposium}.\hskip 1em plus 0.5em minus 0.4em\relax {IEEE}.
  10.1109/ipdps.2005.335. [Online]. Available:
  \url{https://doi.org/10.1109/ipdps.2005.335}
\BIBentrySTDinterwordspacing

\bibitem{heffte}
\BIBentryALTinterwordspacing
A.~Ayala, S.~Tomov, A.~Haidar, and J.~Dongarra, ``{heFFTe}: Highly efficient
  {FFT} for exascale,'' in \emph{Lecture Notes in Computer Science}.\hskip 1em
  plus 0.5em minus 0.4em\relax Springer International Publishing, 2020, pp.
  262--275. [Online]. Available:
  \url{https://doi.org/10.1007/978-3-030-50371-0_19}
\BIBentrySTDinterwordspacing

\bibitem{gholami2016accfft}
\BIBentryALTinterwordspacing
A.~Gholami, J.~Hill, D.~Malhotra, and G.~Biros, ``{AccFFT}: A library for
  distributed-memory {FFT} on {CPU} and {GPU} architectures,''
  \emph{arXiv:1506.07933}, 2016. [Online]. Available:
  \href{https://arxiv.org/abs/1506.07933}{1506.07933}
\BIBentrySTDinterwordspacing

\bibitem{multi_process_fft}
\BIBentryALTinterwordspacing
A.~Ayala, S.~Tomov, M.~Stoyanov, A.~Haidar, and J.~Dongarra, ``Accelerating
  multi - process communication for parallel 3-d {FFT},'' in \emph{2021
  Workshop on Exascale {MPI} ({ExaMPI})}.\hskip 1em plus 0.5em minus
  0.4em\relax {IEEE}, Nov. 2021.  10.1109/exampi54564.2021.00011. [Online].
  Available: \url{https://doi.org/10.1109/exampi54564.2021.00011}
\BIBentrySTDinterwordspacing

\bibitem{ayala2021interim}
\BIBentryALTinterwordspacing
A.~Ayala, S.~Tomov, P.~Luszczek, S.~Cayrols, G.~Ragghianti, and J.~Dongarra,
  ``Interim report on benchmarking {FFT} libraries on high performance
  systems.'' [Online]. Available:
  \url{https://icl.utk.edu/publications/interim-report-benchmarking-fft-libraries-high-performance-systems}
\BIBentrySTDinterwordspacing

\bibitem{Qiskit}
{Qiskit contributors}, ``Qiskit: An open-source framework for quantum
  computing,'' 2023.

\bibitem{sharma2022openqaoa}
\BIBentryALTinterwordspacing
V.~Sharma, N.~S.~B. Saharan, S.-H. Chiew, E.~I.~R. Chiacchio, L.~Disilvestro,
  T.~F. Demarie, and E.~Munro, ``{OpenQAOA} -- an {SDK} for {QAOA},''
  \emph{arXiv:2210.08695}, 2022. [Online]. Available:
  \href{https://arxiv.org/abs/2210.08695}{2210.08695}
\BIBentrySTDinterwordspacing

\bibitem{Shaydulin2021QAOAKit}
\BIBentryALTinterwordspacing
R.~Shaydulin, K.~Marwaha, J.~Wurtz, and P.~C. Lotshaw, ``{QAOAKit}: A toolkit
  for reproducible study, application, and verification of the {QAOA},'' in
  \emph{2021 {IEEE}/{ACM} Second International Workshop on Quantum Computing
  Software ({QCS})}.\hskip 1em plus 0.5em minus 0.4em\relax {IEEE}, Nov. 2021.
  10.1109/qcs54837.2021.00011. [Online]. Available:
  \url{https://doi.org/10.1109/qcs54837.2021.00011}
\BIBentrySTDinterwordspacing

\bibitem{cuquantum}
``{cuQuantum SDK},'' \url{https://developer.nvidia.com/cuquantum-sdk},
  accessed: 2023-07-20.

\bibitem{QTensor}
``{QTensor simulator on GitHub},'' \url{https://github.com/danlkv/QTensor},
  accessed: 2023-07-20.

\bibitem{Schutski_adaptive}
\BIBentryALTinterwordspacing
R.~Schutski, D.~Lykov, and I.~Oseledets, ``Adaptive algorithm for quantum
  circuit simulation,'' \emph{Phys. Rev. A}, vol. 101, p. 042335, Apr 2020.
  10.1103/PhysRevA.101.042335. [Online]. Available:
  \url{https://link.aps.org/doi/10.1103/PhysRevA.101.042335}
\BIBentrySTDinterwordspacing

\bibitem{Lykov_diagonal_gates}
\BIBentryALTinterwordspacing
D.~Lykov and Y.~Alexeev, ``Importance of diagonal gates in tensor network
  simulations,'' in \emph{2021 IEEE Computer Society Annual Symposium on VLSI
  (ISVLSI)}, 2021.  10.1109/ISVLSI51109.2021.00088 pp. 447--452. [Online].
  Available: \url{https://doi.org/10.1109/ISVLSI51109.2021.00088}
\BIBentrySTDinterwordspacing

\bibitem{bayraktar2023cuquantum}
\BIBentryALTinterwordspacing
H.~Bayraktar, A.~Charara, D.~Clark, S.~Cohen, T.~Costa, Y.-L.~L. Fang, Y.~Gao,
  J.~Guan, J.~Gunnels, A.~Haidar, A.~Hehn, M.~Hohnerbach, M.~Jones, T.~Lubowe,
  D.~Lyakh, S.~Morino, P.~Springer, S.~Stanwyck, I.~Terentyev, S.~Varadhan,
  J.~Wong, and T.~Yamaguchi, ``{cuQuantum} {SDK}: A high-performance library
  for accelerating quantum science,'' \emph{arXiv:2308.01999}, 2023. [Online].
  Available: \href{https://arxiv.org/abs/2308.01999}{2308.01999}
\BIBentrySTDinterwordspacing

\bibitem{Dahi2023ASO}
\BIBentryALTinterwordspacing
Z.~A. E.~M. Dahi, E.~Alba, R.~Gil-Merino, F.~Chicano, and G.~Luque, ``A survey
  on quantum computer simulators,'' 2023. [Online]. Available:
  \url{https://api.semanticscholar.org/CorpusID:259257304}
\BIBentrySTDinterwordspacing

\bibitem{Gray_2021}
\BIBentryALTinterwordspacing
J.~Gray and S.~Kourtis, ``Hyper-optimized tensor network contraction,''
  \emph{Quantum}, vol.~5, p. 410, mar 2021.  10.22331/q-2021-03-15-410.
  [Online]. Available: \url{https://doi.org/10.22331%2Fq-2021-03-15-410}
\BIBentrySTDinterwordspacing

\bibitem{zulehner2019advanced_decision_diagrams}
\BIBentryALTinterwordspacing
A.~Zulehner and R.~Wille, ``Advanced simulation of quantum computations,''
  \emph{{IEEE} Transactions on Computer-Aided Design of Integrated Circuits and
  Systems}, vol.~38, no.~5, pp. 848--859, May 2019.  10.1109/tcad.2018.2834427.
  [Online]. Available: \url{https://doi.org/10.1109/tcad.2018.2834427}
\BIBentrySTDinterwordspacing

\bibitem{burgholzer2020advanced}
\BIBentryALTinterwordspacing
L.~Burgholzer and R.~Wille, ``Advanced equivalence checking for quantum
  circuits,'' \emph{{IEEE} Transactions on Computer-Aided Design of Integrated
  Circuits and Systems}, vol.~40, no.~9, pp. 1810--1824, Sep. 2021.
  10.1109/tcad.2020.3032630. [Online]. Available:
  \url{https://doi.org/10.1109/tcad.2020.3032630}
\BIBentrySTDinterwordspacing

\bibitem{lykov_step_dependent_qtensor}
\BIBentryALTinterwordspacing
D.~Lykov, R.~Schutski, A.~Galda, V.~Vinokur, and Y.~Alexeev, ``Tensor network
  quantum simulator with step-dependent parallelization,'' in \emph{2022 {IEEE}
  International Conference on Quantum Computing and Engineering ({QCE})}.\hskip
  1em plus 0.5em minus 0.4em\relax {IEEE}, Sep. 2022.
  10.1109/qce53715.2022.00081. [Online]. Available:
  \url{https://doi.org/10.1109/qce53715.2022.00081}
\BIBentrySTDinterwordspacing

\bibitem{cameron_contraction_trees}
\BIBentryALTinterwordspacing
C.~Ibrahim, D.~Lykov, Z.~He, Y.~Alexeev, and I.~Safro, ``Constructing optimal
  contraction trees for tensor network quantum circuit simulation,'' in
  \emph{2022 {IEEE} High Performance Extreme Computing Conference
  ({HPEC})}.\hskip 1em plus 0.5em minus 0.4em\relax {IEEE}, Sep. 2022.
  10.1109/hpec55821.2022.9926353. [Online]. Available:
  \url{https://doi.org/10.1109/hpec55821.2022.9926353}
\BIBentrySTDinterwordspacing

\bibitem{khakhulin2020learning}
\BIBentryALTinterwordspacing
T.~Khakhulin, R.~Schutski, and I.~Oseledets, ``Learning elimination ordering
  for tree decomposition problem,'' in \emph{Learning Meets Combinatorial
  Algorithms at NeurIPS2020}, 2020. [Online]. Available:
  \url{https://openreview.net/forum?id=aZ7wAnYs9v1}
\BIBentrySTDinterwordspacing

\bibitem{Meirom2022OptimizingTN}
E.~Meirom, H.~Maron, S.~Mannor, and G.~Chechik, ``Optimizing tensor network
  contraction using reinforcement learning,'' in \emph{International Conference
  on Machine Learning}.\hskip 1em plus 0.5em minus 0.4em\relax PMLR, 2022, pp.
  15\,278--15\,292.

\bibitem{gray2022hyperoptimized}
\BIBentryALTinterwordspacing
J.~Gray and S.~Kourtis, ``Hyper-optimized tensor network contraction,'' p. 410,
  Mar. 2021. [Online]. Available:
  \url{https://doi.org/10.22331/q-2021-03-15-410}
\BIBentrySTDinterwordspacing

\bibitem{Contracting_arbitrary_tensor_networks}
\BIBentryALTinterwordspacing
F.~Pan, P.~Zhou, S.~Li, and P.~Zhang, ``Contracting arbitrary tensor networks:
  General approximate algorithm and applications in graphical models and
  quantum circuit simulations,'' \emph{Phys. Rev. Lett.}, vol. 125, p. 060503,
  Aug 2020.  10.1103/PhysRevLett.125.060503. [Online]. Available:
  \url{https://link.aps.org/doi/10.1103/PhysRevLett.125.060503}
\BIBentrySTDinterwordspacing

\bibitem{full_state_compression}
\BIBentryALTinterwordspacing
X.-C. Wu, S.~Di, E.~M. Dasgupta, F.~Cappello, H.~Finkel, Y.~Alexeev, and F.~T.
  Chong, ``Full-state quantum circuit simulation by using data compression,''
  in \emph{Proceedings of the International Conference for High Performance
  Computing, Networking, Storage and Analysis}, ser. SC '19, 2019.
  10.1145/3295500.3356155. [Online]. Available:
  \url{https://doi.org/10.1145/3295500.3356155}
\BIBentrySTDinterwordspacing

\bibitem{cirq_approximate}
\BIBentryALTinterwordspacing
S.~V. Isakov, D.~Kafri, O.~Martin, C.~V. Heidweiller, W.~Mruczkiewicz, M.~P.
  Harrigan, N.~C. Rubin, R.~Thomson, M.~Broughton, K.~Kissell, E.~Peters,
  E.~Gustafson, A.~C.~Y. Li, H.~Lamm, G.~Perdue, A.~K. Ho, D.~Strain, and
  S.~Boixo, ``Simulations of quantum circuits with approximate noise using qsim
  and {C}irq,'' \emph{arXiv:2111.02396}, 2021. [Online]. Available:
  \href{https://arxiv.org/abs/2111.02396}{2111.02396}
\BIBentrySTDinterwordspacing

\bibitem{tensorflow_quantum}
\BIBentryALTinterwordspacing
M.~Broughton, G.~Verdon, T.~McCourt, A.~J. Martinez, J.~H. Yoo, S.~V. Isakov,
  P.~Massey, R.~Halavati, M.~Y. Niu, A.~Zlokapa, E.~Peters, O.~Lockwood,
  A.~Skolik, S.~Jerbi, V.~Dunjko, M.~Leib, M.~Streif, D.~Von~Dollen, H.~Chen,
  S.~Cao, R.~Wiersema, H.-Y. Huang, J.~R. McClean, R.~Babbush, S.~Boixo,
  D.~Bacon, A.~K. Ho, H.~Neven, and M.~Mohseni, ``Tensorflow quantum: A
  software framework for quantum machine learning,'' \emph{arXiv:2003.02989},
  2020. [Online]. Available:
  \href{https://arxiv.org/abs/2003.02989}{2003.02989}
\BIBentrySTDinterwordspacing

\bibitem{Efthymiou2022}
\BIBentryALTinterwordspacing
S.~Efthymiou, M.~Lazzarin, A.~Pasquale, and S.~Carrazza, ``Quantum simulation
  with just-in-time compilation,'' \emph{Quantum}, vol.~6, p. 814, Sep. 2022.
  10.22331/q-2022-09-22-814. [Online]. Available:
  \url{https://doi.org/10.22331/q-2022-09-22-814}
\BIBentrySTDinterwordspacing

\bibitem{Smelyanskiy2016qHiPSTERTQ}
\BIBentryALTinterwordspacing
M.~Smelyanskiy, N.~P.~D. Sawaya, and A.~Aspuru-Guzik, ``{qHiPSTER}: The quantum
  high performance software testing environment,'' \emph{arXiv:1601.07195},
  2016. [Online]. Available:
  \href{https://arxiv.org/abs/1601.07195}{1601.07195}
\BIBentrySTDinterwordspacing

\bibitem{Shaydulin2021}
\BIBentryALTinterwordspacing
R.~Shaydulin, S.~Hadfield, T.~Hogg, and I.~Safro, ``Classical symmetries and
  the quantum approximate optimization algorithm,'' \emph{Quantum Information
  Processing}, vol.~20, no.~11, Oct. 2021.  10.1007/s11128-021-03298-4.
  [Online]. Available: \url{https://doi.org/10.1007/s11128-021-03298-4}
\BIBentrySTDinterwordspacing

\bibitem{Shaydulin2021TQE}
\BIBentryALTinterwordspacing
R.~Shaydulin and S.~M. Wild, ``Exploiting symmetry reduces the cost of training
  {QAOA},'' \emph{{IEEE} Transactions on Quantum Engineering}, vol.~2, pp.
  1--9, 2021.  10.1109/tqe.2021.3066275. [Online]. Available:
  \url{https://doi.org/10.1109/tqe.2021.3066275}
\BIBentrySTDinterwordspacing

\bibitem{2211.09270}
\BIBentryALTinterwordspacing
J.~Sud, S.~Hadfield, E.~Rieffel, N.~Tubman, and T.~Hogg, ``A parameter setting
  heuristic for the quantum alternating operator ansatz,''
  \emph{arXiv:2211.09270}, 2022. [Online]. Available:
  \href{https://arxiv.org/abs/2211.09270}{2211.09270}
\BIBentrySTDinterwordspacing

\bibitem{TQA_github}
``Simulation code for trotterized quantum annealing on {G}it{H}ub,''
  \url{https://github.com/shsack/TQA-init.-for-QAOA/blob/59aaca45c382bc0b8ec93b0810a8d5ce45c2f28d/TQA_QAOA.ipynb},
  accessed: 2023-09-12.

\bibitem{Sack2021quantumannealing}
\BIBentryALTinterwordspacing
S.~H. Sack and M.~Serbyn, ``Quantum annealing initialization of the quantum
  approximate optimization algorithm,'' \emph{{Quantum}}, vol.~5, p. 491, Jul.
  2021.  10.22331/q-2021-07-01-491. [Online]. Available:
  \url{https://doi.org/10.22331/q-2021-07-01-491}
\BIBentrySTDinterwordspacing

\end{thebibliography}

\section*{Disclaimer}
This paper was prepared for informational purposes with contributions from the Global Technology Applied Research center of JPMorgan Chase \& Co. This paper is not a product of the Research Department of JPMorgan Chase \& Co. or its affiliates. Neither JPMorgan Chase \& Co. nor any of its affiliates makes any explicit or implied representation or warranty and none of them accept any liability in connection with this position paper, including, without limitation, with respect to the completeness, accuracy, or reliability of the information contained herein and the potential legal, compliance, tax, or accounting effects thereof. This document is not intended as investment research or investment advice, or as a recommendation, offer, or solicitation for the purchase or sale of any security, financial instrument, financial product or service, or to be used in any way for evaluating the merits of participating in any transaction.

The submitted manuscript includes contributions from 
UChicago Argonne, LLC, Operator of Argonne National Laboratory (``Argonne'').
Argonne, a U.S.\ Department of Energy Office of Science laboratory, is operated
under Contract No.\ DE-AC02-06CH11357.  The U.S.\ Government retains for itself,
and others acting on its behalf, a paid-up nonexclusive, irrevocable worldwide
license in said article to reproduce, prepare derivative works, distribute
copies to the public, and perform publicly and display publicly, by or on
behalf of the Government.  The Department of Energy will provide public access
to these results of federally sponsored research in accordance with the DOE
Public Access Plan \url{http://energy.gov/downloads/doe-public-access-plan}.

\end{document}